\documentclass[9pt,twocolumn,twoside,lineno]{pnas-new}
\templatetype{pnasresearcharticle} 

\title{SOS -- Self-Organization for Survival: Introducing \textit{fairness} in emergency communication to save
lives}

\author[a,1]{Indushree Banerjee}
\author[a]{Martijn Warnier}
\author[a]{Frances M.T.\ Brazier}
\author[b,1]{Dirk Helbing}

\affil[a]{Systems Engineering and Simulation, Faculty of Technology Policy and Management, Delft University of Technology}
\affil[b]{Computational Social Science, ETH Zurich}
\leadauthor{Banerjee} 

\significancestatement{In an unsustainable world, the frequency and severity of disasters is expected to increase.  Disasters often damage telecommunication infrastructure and cause electricity blackouts that prevent citizens from recharging their phones. Most current emergency communication apps drain energy without consideration of battery charge in phones. This puts people at risk of losing their ability to communicate when they need it most. In this article, we focus on these considerations when designing an emergency communication network for participatory resilience. We demonstrate how a value-sensitive design can make sure that a maximum number of people is able to communicate for an extended period of time. Empowering citizens is important because they are often the first to respond, especially when sites are cut off and public rescue efforts start with a delay.}

\authorcontributions{ I.B., M.W., F.B. and D.H. designed the research, discussed the results, outlined the paper.
\par I.B. developed the protocol, implemented the simulation and modeling and conducted the experiments in close interaction with M.W.
\par I.B., M.W., F.B. and D.H. wrote and edited the paper.}
\authordeclaration{The authors declare no conflict of interest.}
\correspondingauthor{\textsuperscript{1}To whom correspondence should be addressed. E-mail:i.banerjee@tudelft.nl; \and dirk.helbing@gess.ethz.ch}


\begin{abstract}
Communication is crucial when disasters isolate communities of people and traditional rescue is delayed. Such delays force citizens to be first responders and form small rescue teams. Rescue teams require reliable communication, particularly in the first 72 hours, which is challenging due to damaged infrastructure and electrical blackouts. We design a peer-to-peer communication network that meets these challenges. 
We introduce the concept of participatory fairness: equal communication opportunities for all citizens regardless of initial inequality in phone battery charge. Our value-sensitive design approach achieves an even battery charge distribution across phones over time and enables citizens to communicate over 72 hours. We apply the fairness principle to communication in an adapted standard Barabasi-Albert model of a scale-free network that automatically (i) assigns high-battery phones as hubs, (ii) adapts the network topology to the spatio-temporal battery charge distribution, and (iii) self-organizes to remain robust and reliable when links fail or phones leave the network. 
While the Barabasi-Albert model has become a widespread descriptive model, we demonstrate its use as a design principle to meet values such as fairness and systemic efficiency. Our results demonstrate that, compared to a generic peer-to-peer mesh network, the new protocol achieves (i) a longer network lifetime, (ii) an adaptive information flow, (iii) a fair distribution of battery charge, and (iv) higher participation rates. Hence, our protocol, Self-Organization for Survival ('SOS'), provides fair communication opportunities to all citizens during a disaster through self-organization. SOS enables participatory resilience and sustainability, empowering citizens to communicate when they need it most.
\end{abstract}

\dates{This manuscript was compiled on \today}
\doi{\url{www.pnas.org/cgi/doi/10.1073/pnas.XXXXXXXXXX}}

\begin{document}

\maketitle
\thispagestyle{firststyle}
\ifthenelse{\boolean{shortarticle}}{\ifthenelse{\boolean{singlecolumn}}{\abscontentformatted}{\abscontent}}{}

\begin{quote}
    ``How effective are existing innovative ways to share data in humanitarian settings, such as mesh networks, bluetooth technology, microwave technology and peer-to-peer networks? What other novel strategies exist?''
From: \textit{``Grand Challenges of Humanitarian Aid''},\\ 
Nature, 2018 \cite{daar2018grand}.
\end{quote}
\dropcap{W}e are living in a world in which the frequency and severity of natural disasters -- causing deaths and displacements -- are steadily increasing \cite{yeeles2018unequal}. This is also because of tipping points \cite{dakos2008slowing} and cascading effects \cite{scheffer2001catastrophic,duan2019universal} in complex anthropogenic systems \cite{steffen2018trajectories}. The aftermath of Hurricane Katrina, the Nepal earthquake, and the Indian Ocean tsunami has shown that delays in rescue operations lead to the loss of additional human lives \cite{daar2018grand}. The number of casualties could be reduced if interventions such as preliminary first aid and basic support were provided during the first 72 hours following a disaster, called the ``Golden Period'' \cite{kohn2012personal}. However, mobilizing rescue operations and professional help for disaster recovery takes time \cite{PhysRevE.75.056107, kohn2012personal}. It is, therefore, crucial that citizens are provided with tools that enable participatory resilience and sustainability, allowing them to help themselves and support each other \cite{Stephenson2007, Aldrich2011,kohn2012personal}. In this article, we offer a solution to one of the grand challenges of humanitarian aid \cite{daar2018grand}. This promises affected communities to have extended and increased access to communication via a peer-to-peer communication network that is designed for fair participation. 
\par Reliable communication in a dynamically changing environment \cite{winerman2009crisis} is challenging \cite{manoj2007communication}. The challenge to stay connected during disasters is increased by (i) failure of telecommunication infrastructures due to damage, and (ii) limited battery charge in phones due to power blackouts\cite{Cetinkaya2010,Little2002, Bashan2013, DeLaRee2005,Chang2007,noam1996kobe}. Power grids and mobile telecommunication are highly interdependent, so the failure of one has a cascading effect on the other \cite{helbing2013globally,vespignani2010fragility}. For example, 8000 mobile base stations immediately failed in Japan on March 2011 after the tsunami. This number doubled by the following day as backup power was exhausted, which led to 85\% of mobile communication breaking down during this time \cite{kobayashi2014experience}. Hurricane Katrina damaged three million telephone land-lines, disabling numerous 911 call centers. With approximately 2,000 cell sites uprooted, and limited locations to charge phones due to power outages, many wireless phones were not reachable \cite{Cetinkaya2010}.

\par Various smartphone applications have been developed that are promoted as facilitating emergency communication \cite{Floreano2015} and supporting participatory disaster response \cite{Helsloot2004}. These applications utilize wireless capabilities of end-user devices such as Bluetooth and Wi-Fi to exchange messages peer-to-peer, forming an ``ad hoc'' communication network on-the-fly \cite {Legendre2011,bruno2005mesh}. 

\par In these applications, there is typically a direct point-to-point connection between all phones that are in transmission range of each other. If sender and receiver are not in transmission range, the message is relayed by other people’s phones. This connection pattern is termed a mesh topology \cite{bruno2005mesh}.  

\par A mesh topology is the standard connection pattern for existing generic applications such as FireChat, ServalMesh \cite{gardner2013rational}, BATMAN \cite{johnson2008simple} and Twimight \cite{Legendre2011}. However, these applications were not specifically designed for the purpose of communication after a disaster. Extended disaster response requires different design principles. 
\par 
This computational social science \cite{lazer2009computational} paper goes beyond the current state of the art in three ways: First, we introduce a value-sensitive design approach for communication networks. Second, while the preferential attachment approach has become a widespread descriptive model, for example, for the structure of the Internet, protein-protein networks, or scientific citation networks, we demonstrate how it can be used as design principle to improve the performance of communication networks under stress. Third, we manage to boost participatory resilience by introducing participatory fairness, thereby benefitting disaster-struck communities over extended periods of time, when they need communication to survive and help each other.
\par
In the following, we discuss the implementation and advantages of a value-sensitive design called ``Self-Organization for Survival'' (``SOS''), which is specifically made for disaster scenarios. It can benefit individuals and promote collective behavior based on local interactions \cite{ballerini2008interaction}. Compared to a typical mesh network, the design of SOS ensures that phones with different battery charge have the opportunity to communicate for 72 hours without recharging. It does so by considering the additional value of ``fairness''. 

\section{Methodological approach: Design for Values }
``Design for values'' is an approach to include values such as autonomy, fairness, usability, privacy, or democracy in the design and operation of technology \cite{van2007ict,friedman1997software}.
\par 
Existing applications establishing an infrastructure-less mesh network are implicitly or explicitly designed for citizen-based communication.  They provide autonomy from backbone infrastructure and reliability in communication.  To facilitate collaboration and communication of a community of people during an unexpected disaster, other factors need to be considered, too. These include the unavailability of charging opportunities and the amount of time phones must be able to communicate, while the network needs to continuously adapt to a changing environment.
\par 
If an application is not designed with these factors in mind, biases may arise that could compromise the outcome in three ways \cite{friedman1996bias}: through preexisting biases, technical biases, and/or emergent biases.
Preexisting biases are rooted in the fabrics of society. Technical biases refer to technical constraints or issues. Emergent biases result from usage, which may depend on the context. Today's ad hoc networks have technical biases due to technical limitations and emergent biases due to lack of consideration. Technical restrictions such as lack of charging facilities and limited resources of phones may also imply emergent biases such as the disparity of communication opportunities. We address the issue of participation disparity in the following.

\par 
Our paper pursues a design-for-values approach to reduce technical and emergent biases in a dynamically changing context. We reorganise information traffic in an adaptive manner to achieve \emph{fairness}. SOS enables and maintains the participation of practically \emph{all} phones, i.e. it provides equal communication opportunities for all citizens regardless of initial inequality in phone battery charge. Overcoming inequality serves to keep the social fabric functional under stress, for example, during crisis and disasters.  

\begin{figure*}
\centering
\includegraphics[width=0.8\textwidth]{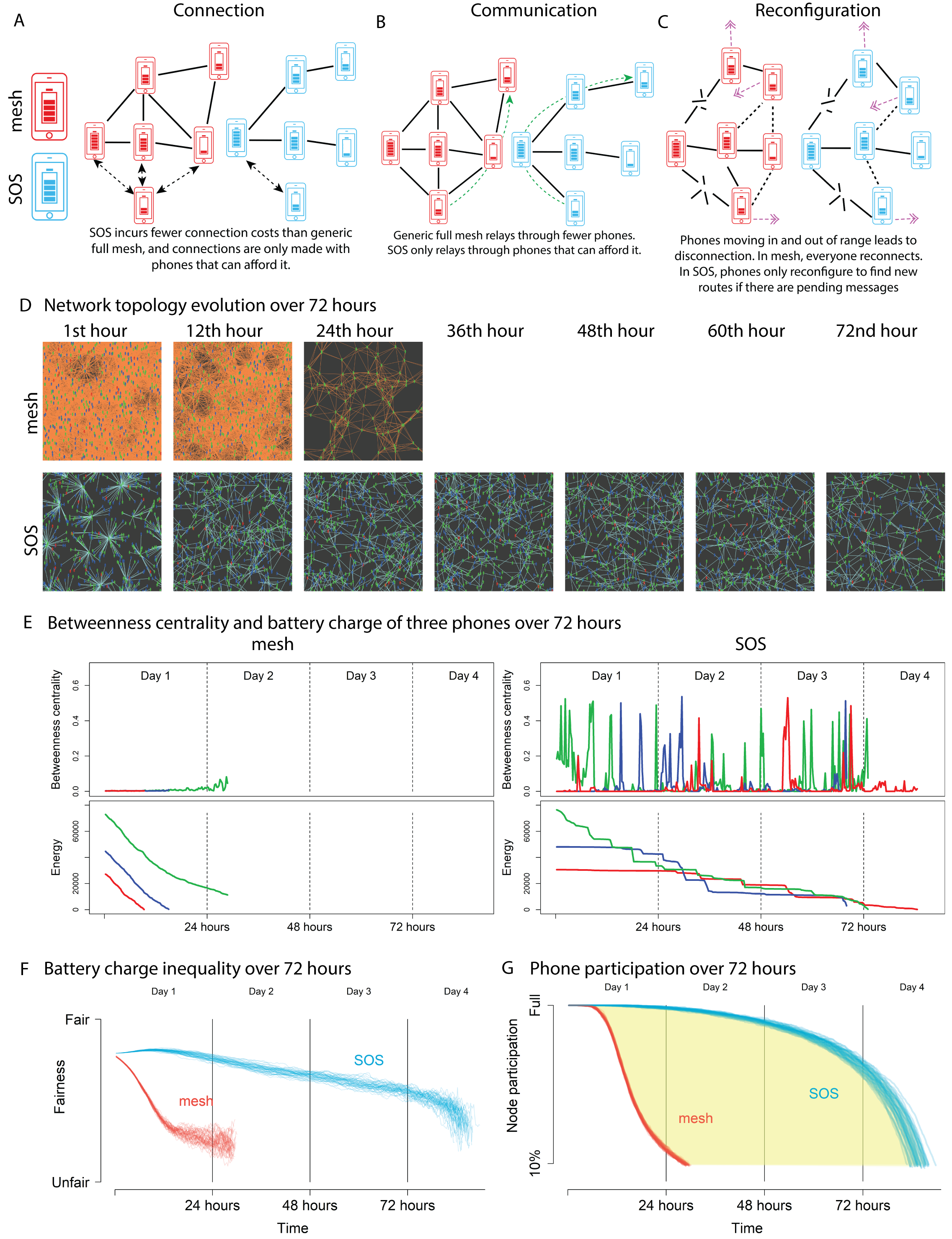}
\caption{
(A) Differences in connection, (B) communication and (C) reconfiguration patterns between a generic mesh protocol (red) and SOS (blue). (D)-(G) Results of simulations with 500 phones, sending and receiving 1 message per phone every 15 minutes for the generic mesh and SOS protocols. (B) shows screenshots of the simulation for different time points. (D) Top: Formation and evolution of the mesh topology. Bottom: Illustration for a non-cyclic preferentially attached network (SOS). The ad hoc mesh network runs just longer than a day. SOS runs for the entire duration of 72 hours. (E) Development of Battery charge and Betweenness Centrality for a selection of three typical phones (red: low initial battery charge; blue: average initial battery charge; green: high initial battery charge). Left: mesh network; right: SOS. In the mesh network, every phone has the same Betweenness Centrality. In SOS, the Betweenness Centrality fluctuates, with green starting as a central hub; a role which is later taken over by blue and then red, as the relative battery charge changes. This ensures that all phones can equally participate in communication for an extended time period. (F) Development of battery charge inequality (Gini coefficient\cite{druckman2008measuring}) over 72 hours for the mesh network (red) and for SOS (blue). (G) Phone participation over 72 hours for the mesh network (red) and for SOS (blue), with the difference in yellow.}
\label{fig:illutration}
\end{figure*}

\subsection{SOS: Designing a \textit{fair} protocol}
A phone loses battery charge when connecting to another phone, sending a message, receiving a message, or relaying a message. In mesh networks, direct peer-to-peer connections result in fewer relays, thus they deliver a message more quickly. It appears that this is energy-efficient, because costs associated with relaying messages through multiple hops are reduced. However, in a traditional mesh topology, connection costs are high as every phone in the transmission range connects (Fig. 1A; red phones). When two phones connect, both phones lose battery charge, making it an expensive procedure.  

\par SOS avoids redundant connections to reduce energy costs. SOS uses local knowledge exchange to form a scale-free network based on preferential attachment \cite{barabasi1999emergence, jeong2003measuring, vazquez2003growing,amaral2004emergence}. The preferential attachment approach consists of two rules: (i) only connect to the highest battery charge phone in the transmission range; (ii) prevent loops by not connecting to a phone in range if already connected via intermediary phones. To execute these rules, every participating phone needs to exchange local knowledge that consists of (i) a phone identifier, (ii) the battery charge left in the phone, and (iii) a network identifier (see SI for further details).  

\par Based on this information, every phone decides on one other phone with which to connect. When two phones connect they equalize their network identifier (see SI for Algorithms and Pseudocode). This leads to the formation of a preferentially attached non-cyclic scale-free network (see Fig. 1A; blue phones). Note that the topology that results from the presented algorithm can be combined with \emph{any} routing protocol for energy efficiency and adds topology control protocols to existing state-of-the-art mechanisms for peer-to-peer communication. In mesh networks, messages may be routed via low battery charge phones. SOS, in contrast, avoids this. To maintain fairness, high battery charge phones automatically become hubs as long as they have higher battery charge than their neighbours (see Fig. 1B; blue phones). 

\par Event-driven reconfiguration is triggered when a phone tries to send a message but has no route to the receiver (see Fig. 1C (blue phones) and SI on Algorithm). The message is saved as a pending message in the phone. If the phone is completely disconnected, it tries to find new connections in its range. Otherwise the phone requests its connected phones to update their local knowledge and form new connections. Local knowledge is exchanged with new phones that are now in range. Each phone follows the connection rule, ensuring that low battery charge phones detect new higher battery charge phones and connect to them. This creates a bigger connected network with highest battery charge phones as hubs and new routes for communication. The advantage of using local knowledge to form scale-free loop-free preferentially attached network is the inherent robustness in the connectivity distribution and the average shortest path length \cite{PhysRevE.69.037103} 

\par Further explanations of functional and non-functional requirements, subsequent design choices and their implications as well as the pseudo-code of the SOS protocol are presented in the SI. 


\subsection{Modeling and simulation}
We use an agent-based modeling \cite{bonabeau2002agent} approach to compare SOS to a generic mesh network. For the results of Fig. 1, we simulate a torus-shaped world of 25x25 units with 500 people with phones, with normally distributed battery charge (see SI for more details). In the model, people with phones move at a constant speed in a random walk (which simplifies typically observed mobility patterns \cite{deville2016scaling}). For simplicity, each phone sends one message every fifteen minutes to a randomly chosen other phone (even though real message frequency is not this homogeneous \cite{wu2010evidence}). The transmission range is assumed to be homogeneous at 5 units. The loss of battery charge associated with sending, receiving, relaying, connecting and reconfiguring is specified according to real Bluetooth Low Energy battery costs \cite{TexasInstruments}. 

\par For the results of Fig. 2, settings were the same as above, with the exception of the number of people and message frequency. The number of people was systematically varied from 100 to 800 to estimate the effect of population density on longevity. Message frequency was varied between one and ten messages sent every fifteen minutes, to estimate the effect of the amount of data traffic on longevity.


\section{Results}
\par The performance of SOS was compared to a mesh communication topology through modeling and simulation in terms of longevity, traffic adaptivity, battery charge inequality, and phone participation (see the SI for details of the experimental setup).

\subsection{SOS lasts for 72 hours, traditional mesh for little more than 24 hours}
 Fig. 1D (and movie S1 in SI) shows the development of the topology over 72 hours for the mesh (top) and SOS (bottom) communication topologies.  
 The mesh topology is tightly coupled with peer-to-peer connections between phones in each others' transmission range in the first hour. This continues and results in a crowded topology for 24 hours. Despite phones moving in and out of range, there is no noticeable change in topology. This is due to the "connect to all in range" characteristics of mesh.

 \par The SOS topology shows a non-cyclic preferentially attached network. This is most clearly visible in the 1st hour. Some phones (with a high battery charge) have many phones connected to them, acting as hubs. Others (with a low battery charge) have one connection and lie on the edge of the network. Initial high battery charge phones later on take less central positions in the network, when other high battery charge phones take over as hubs. Over time, this results in a network with a more even energy distribution among phones. 

\subsection{SOS adapts the traffic distribution to spare low-energy phones}
Fig. 1E (and movie S2 and S3 in SI) shows how the adaptive mechanism of SOS affects the consumption of battery charge over time and the Betweenness Centrality. Betweenness Centrality measures the importance of a phone for passing information. Higher Betweenness Centrality shows that a particular phone is more centrally placed in the emerging network, which means maximum data traffic passes through this phone. 

\par For mesh networks (see left of Fig. 1C), the battery charge rank of phones is stable over time. All phones have almost the same Betweenness Centrality with a slight variation at the end towards 0.14, and all phones lose battery charge linearly over time. This creates a discriminatory bias against people with phones that happened to have a low initial battery charge (such as the red phone). They are disconnected earlier, limiting their communication opportunities in favour of people with a higher battery charge (green). 

\par SOS automatically assigns high-energy phones as hubs and monitors the spatio-temporal energy distribution to adapt the network topology. This mechanism prevents selfish behaviour and promotes altruism. This is reflected in the changing Betweenness Centrality. For the SOS protocol, the figure on the right of Fig. 1C shows how the role of phones changes over time.  The green phone initially plays a central role (with a Betweenness Centrality of 0.52 after two hours), because it has a high level of battery charge, while other phones are spared. After some time, the blue phone becomes a hub (with a Betweenness Centrality of 0.54, peaks between hours 25 and 38, and again towards the end). After that, there is a period where the red phone becomes a hub (with a Betweenness Centrality of 0.53). 

\par The red phone starts with the lowest battery charge and is the first to disconnect in the mesh network. In SOS, the red phone is spared from relaying messages, allowing it to stay connected for as long as the green phone with the highest initial battery charge. This illustrates how the topology adapts to the spatio-temporal situation of energy availability. The phones keep changing with regard to the load and traffic to spare the lower battery charge phones, such that fairness is achieved. 

\subsection{SOS distributes energy more fairly over phones than traditional mesh}
The fairness of the phone battery charge distribution is calculated here with the Gini coefficient. Fig. 1F shows the Gini coefficient over time, for SOS (in blue) and mesh (in red). The Gini coefficient is typically used to study inequality, e.g.\ of income or resources \cite{druckman2008measuring}. Its value ranges from 0 to 1 with 0 signifying complete equality (all have the same battery charge) and 1 meaning extreme inequality (one phone has all battery charge). 
\par For the traditional mesh network, inequality increases quickly, with the Gini coefficient increasing from 0.13 to 0.39 within the first 14 hours, then stabilizing at 0.45. 
For SOS, inequality decreases within the first 10 hours, with the Gini coefficient decreasing from 0.13 to 0.11. Then, within the next 62 hours, the Gini coefficient slowly increases to 0.27. 

\subsection{SOS allows more phones to participate for a longer period than mesh}
Fig. 1G (and movie S1 in SI) shows phone participation over time. For mesh, phones almost immediately start to fail with the first phone dropping out after 3 hours. For SOS, the time period during which there are no failing phones is significantly extended, with the first phone dropping out of the network after 13 hours.  Also, a significant improvement in longevity for SOS is immediately obvious, reflected by the considerable horizontal distance between the SOS and mesh curves. For the mesh topology, only 18\% of phones remain connected after 24 hours, whereas for SOS, 99\% of phones are still connected after 24 hours. For SOS, the phone participation is recorded to be 91\% after 48 hours and 62\% after 72 hours.
\par This illustrates a large difference in the energy efficiency between the two protocols. SOS has several advantages: The communication network lasts considerably longer, and the percentage of phones participating in the network is larger at every point in time. The large separation between the two participation curves demonstrates the success of SOS.
\begin{figure}
\includegraphics[width=0.5\textwidth]{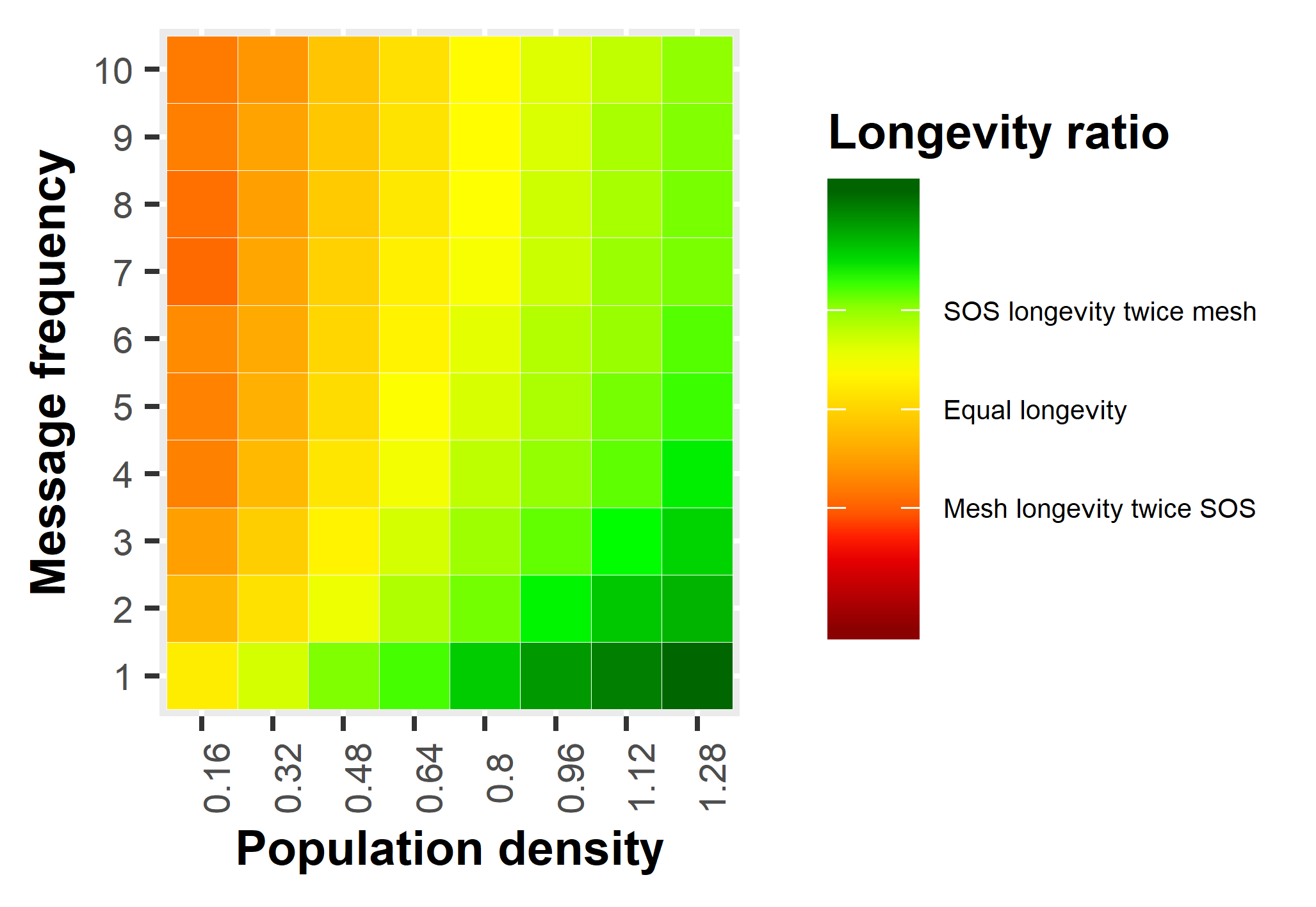}
\caption{Phase diagram of the difference in longevity between mesh and SOS, for varying message frequency and population density. Yellow to dark green indicates an increasing advantage for SOS over mesh. Red to dark red indicates an increasing advantage for mesh over SOS (dark red not featured). Orange indicates that longevity was equal between the two. SOS performs best when population densities are higher (towards the right) or message frequencies are lower (towards the bottom). Mesh performs best when population densities are low and message frequency is high.}
\label{fig:Phase}
\end{figure}

\subsection{The relative advantage of SOS depends on the density and message frequency}
Fig. 2 shows the difference in longevity between mesh and SOS, as a function of message frequency and population density. Message frequency ranged from sending 1 to 10 messages per 15 minutes. Population density is varied by increasing the number of phones in a fixed area. Population density varies from 0.16, representing 100 phones, to 1.28, representing 800 phones.

\par The primary energy expenditure for mesh topology lies in connecting phones. Therefore, the right side of Fig. 2 -- where there are more phones and thus more connections -- shows larger advantages for SOS. The primary energy expenditure for SOS comes from relaying, as routes are longer. Therefore, the lower side of Fig. 2 -- where fewer messages are sent and relayed -- shows larger advantages for SOS. Conversely, the mesh topology performs better in scenarios with low densities and large numbers of messages. 
\par Note that the number of messages for scenarios where the mesh topology outperforms SOS is extreme, with every phone continuously sending multiple messages in each time step. For all other scenarios, SOS outperforms the mesh topology, with the best performance for high phone densities (as in disaster-struck cities) and reasonable volumes of information traffic. 

\section {Discussion and Outlook}
The increased penetration of mobile phones in remote parts of the world has opened avenues for their use to improve situational awareness during disasters \cite{deville2014dynamic,kryvasheyeu2016rapid}. We developed a novel protocol for peer-to-peer communication using a ``design for values'' approach. The value of fairness was identified as particularly important in emergency situations. Our protocol achieves fairness by adapting to the spatio-temporal context of a disaster situation. We use preferential attachment to form a scale-free network, following an adapted standard Barab\'{a}si-Albert model. In this network, phones with high battery charge function as hubs, to facilitate emergency communication for those citizens who are in immediate danger and have little battery charge to spare. This is in contrast with the generic mesh topology that underlies previously proposed emergency communication solutions. These solutions form so many connections that they do not provide the required functionality for nearly as long as required.

\par It seems that recent developments in emergency communications have focused more on introducing infrastructure to disaster-struck areas than on peer-to-peer networks. For example, base stations with Wi-Fi capability may be brought to a disaster area, or unmanned aerial devices can provide connectivity \cite{Floreano2015}. As mentioned above, however, the logistics of disaster response typical implies delays for such solutions, while delays are often deadly. That is why a solution such as SOS is needed, which works over an extended period of time in absence of recharging opportunities. Still, we think that every kind of emergency communication solution has its role to play. Generic mesh itself may transmit messages faster than SOS, because there are fewer hops in between. In situations where batteries can be recharged, generic mesh may therefore be preferable to SOS. Hence, it would be helpful if the communication protocol itself would adjust to the situation at hand. Dynamic decentralized switching between communication protocols is something we are currently looking into.

\par There are two reasons why the development of SOS is increasingly urgent: urbanization and increasing income inequality. The advantages of SOS over generic mesh are greatest when the density of phones is large. This means that it would be especially powerful in urban areas. Currently, over half of the world's population is living in cities, and this proportion is still growing. Also, the population density of cities is growing and so is the disparity of resources \cite{galvani2016human}. This suggests that the value of the SOS approach is increasing, and highly needed to protect the interests of those with phones with little battery charge. 

\par In the simulations that we used to illustrate differences between SOS and the mesh topology, battery charge was randomly distributed. In real life, differences in battery charge may also have structural reasons, as phones with more battery charge are more expensive \cite{coronese2019evidence}. With mesh, those with expensive phones will be able to send messages for longer than inexpensive phones. Multiple inexpensive phones will be relaying their messages and losing the ability to communicate quickly, long before the crucial 72 hours after a disaster are over. With SOS, those with expensive phones will be able to support inexpensive phones' communication opportunities, and strengthen the resilience of the social fabric \cite{shirado2019assortative}. Hence, participation fairness, as achieved by SOS, is especially important in countries with a greater income disparity. 

\par  Citizens, however, are not an interchangeable commodity: Not all citizens will contribute in the same way. Different people have different requirements, but can also offer different skills and contributions \cite{freeman2020social}. Thus, it is important that everyone stays connected. Furthermore, abilities and requirements are not static, as they may change over time. By adapting to changing circumstances, and maximizing the strengths of each, one can empower individual citizens and community resilience, without forgetting those that are in need of support. Our communication protocol does just that. Rather than focusing on equality of workload and maximizing the self-interests of individuals, it is based on fairness in the distribution of workload and promoting altruistic behavior, which benefits the community and individuals alike.

\acknow{This research is part of and funded by TU Delft's PhD school ``Engineering Social Technologies for a Responsible Digital Future''.}

\showacknow{} 

\bibliography{pnas-sample}
\end{document}